\documentclass[11pt]{article}
\usepackage{moriond,epsfig,amsmath}

\def\Journal#1#2#3#4{{#1} {\bf #2}, #3 (#4)}

\def\NPB{{\em Nucl. Phys.} B}
\def\PLB{{\em Phys. Lett.}  B}

\def\PRD{{\em Phys. Rev.} D}

\def\be{\begin{equation}}
\def\ee{\end{equation}}
\def\bea{\begin{eqnarray}}
\def\eea{\end{eqnarray}}

\newcommand{\lesssim}{\stackrel{<}{_{\scriptstyle \sim}}}
\newcommand{\gtrsim}{\stackrel{>}{_{\scriptstyle \sim}}}

\newcommand\TeV{\,\mbox{TeV}}
\newcommand\GeV{\,\mbox{GeV}}
\newcommand\MeV{\,\mbox{MeV}}

\newcommand\mpl{M_{\rm P}}

\begin{document}
\vspace*{4cm}
\title{Constraining the curvaton scenario}

\author{Marieke Postma}

\address{The Abdus Salam International Center for Theoretical Physics, 
Strada Costiera 11, 34100 Trieste, Italy}

\maketitle \abstracts{We analyse the curvaton scenario in the context
of supersymmetry.  Supersymmetric theories contain many scalars, and
therefore many curvaton candidates. To obtain a scale invariant
perturbation spectrum, the curvaton mass should be small during
inflation $m \ll H$. This can be achieved by invoking symmetries,
which suppress the soft masses and non-renormalizable terms in the
potential. Other model-independent constraints on the curvaton model
come from nucleosynthesis, gravitino overproduction, and thermal
damping.  The curvaton can work for masses $m \gtrsim 10^4 \GeV$, and
very small couplings (e.g. $h \lesssim 10^{-6}$ for $m \lesssim 10^8
\GeV$).}

\section{The curvaton scenario}

It is now widely believed that the early universe went through a
period of rapid expansion, called inflation.  In addition to
explaining the homogeneity and isotropy of the observable universe,
inflation can provide the seeds for structure formation.  In the usual
picture quantum fluctuations of the slowly rolling inflaton field
``freeze in'' soon after horizon exit, and become essentially a
classical perturbation which remains constant until the moment of
horizon re-entry.  The resultant perturbations are adiabatic and
Gaussian, in agreement with observations~\cite{cmb}.  Moreover, they
solely depend on the form of the inflaton potential, and are
independent of what goes on between horizon exit an re-entry.  This
makes models of inflation predictive, but also restrictive. The
observed, nearly scale-invariant perturbation spectrum requires very
small coupling constants and/or masses, which renders many models
unnatural.  For this reason it is worthwhile to explore alternative
ways of producing density perturbations.

In the curvaton scenario, the adiabatic perturbations are not
generated by the inflaton field, but instead result from isocurvature
perturbations of some other field --- the {\it curvaton} field.
Adiabatic or curvature perturbations are local perturbations of the
curvature of space-time; isocurvature perturbations on the other hand
do not perturb space-time but correspond to a local perturbation in
the equation of state.  After inflation the isocurvature perturbations
have to be converted into adiabatic ones.  Such a conversion takes
place with the growth of the curvaton energy density compared to the
total energy density in the universe.  This alternative method of
producing adiabatic perturbations was first noted years
ago~\cite{early}, but it did not attract much attention until
recently~\cite{curv}.

The usual implementation of the curvaton scenario is the following. If
the curvaton is light with respect to the Hubble constant during
inflation, it will fluctuate freely, leading to condensate formation.
In the post-inflationary epoch the expansion of the universe acts as a
friction term in the equations of motion, and the field remains
effectively frozen at large field value. This stage ends when the
Hubble constant becomes of the order of the curvaton mass, $H \sim
m_\phi$, at which point the curvaton starts oscillating in the
potential well.  During oscillations, the curvaton acts as
non-relativistic matter, and its energy density red shifts as
$\rho_\phi \propto a^{-3}$ with $a$ the scale factor of the universe.
After inflaton decay, the universe becomes radiation dominated, with
the energy density in radiation red shifting as $\rho_\gamma \propto
a^{-4}$.  Hence, the ratio of curvaton energy density to radiation
energy density grows $\rho_\phi/\rho_\gamma \propto a$, and
isocurvature perturbations are transformed into curvature
perturbations. This conversion halts when the curvaton comes to
dominate the energy density, or if this never happens, when it decays.

\section{Supersymmetry}

It seems natural to try to embed the curvaton scenario within
supersymmetric (SUSY) theories.  SUSY theories contain many flat
directions, i.e., directions in field space along which the scalar
potential vanishes in the supersymmetric limit.  The fields
parametrising these directions can condense during inflation, and are
therefore possible curvaton candidates.  The problem, however, is that
inflation is driven by the non-zero energy density stored in the
inflaton field, and necessarily SUSY is broken during inflation.  One
way to see this, is to note that the inflaton potential is a sum of
$F$ and $D$ terms, and that a non-zero $F$ ($D$)-term does not leave
the SUSY transformation of the quarks (gauginos) invariant.  As a
result, soft mass terms are generated, which are typically of the
order of the Hubble constant.  But this is no good: $m_\phi \sim H$
during inflation leads to a large scale dependence of the produced
perturbations, in conflict with observations.

A way out of this is to invoke symmetries.  If inflation is driven by
$D$-terms, soft mass terms are forbidden by gauge symmetries, and the
problem does not arise.  In no-scale type supergravity, a so-called
Heisenberg symmetry forbids mass terms at tree level, and soft masses
are suppressed by loop factors.  Another possibility is to consider
pseudo-Goldstone bosons, whose mass is protected by approximate global
symmetries.

\section{constraints}

There are several model independent constraints on the curvaton
scenario.  We will discuss them briefly here; see the original paper
for more details~\cite{postma}.

First of all, the curvaton scenario should give rise to the observed
spectrum of density perturbations.  Curvature perturbations ${\mathcal
R}$ of the correct magnitude are obtained for~\cite{cmb,curv}
\begin{equation} 
{\mathcal R} \approx \frac{f}{3 \pi} \frac{H_*}{\phi_*} \approx 5 \times
10^{-5}.
\end{equation}
Here the subscript $*$ denotes the quantity at the time observable
scales leave the horizon, some 60 $e$-folds before the end of
inflation. Further, $f = \rho_\phi / \rho_{\rm tot}$ evaluated at the
time of curvaton decay.  If the curvaton contributes less than 1\% to
the total energy density, i.e., $f < 0.01$, then the perturbations
have an unacceptable large non-Gaussianity. If during inflation
$m_\phi \ll H$ --- which is required to get a nearly scale invariant
perturbation spectrum --- quantum fluctuations of the curvaton grow
until $m_\phi^2 \langle \phi^2 \rangle \sim H^4$, with an
exponentially large coherence length.  We will assume that this sets
the initial curvaton amplitude $\phi_* \sim \sqrt{\langle \phi^2
\rangle}$.  The non-detection of tensor perturbations puts an upper
bound on the Hubble scale during inflation $H_* \lesssim 10^{14}
\GeV$.  Finally, in the curvaton scenario the adiabatic density
perturbations can be accompanied by isocurvature perturbations in the
densities of the various components of the cosmic fluid.  There are
particularly strong bounds on the isocurvature perturbations in cold
dark matter.

In the absence of non-renormalizable terms in the potential the
initial curvaton amplitude can be arbitrarily large, as long as the
curvaton energy density is sub-dominant during inflation.  However, to
avoid a period of inflation driven by the curvaton field, the curvaton
energy density should be still sub-dominant at the onset of curvaton
oscillations.  This restricts the amplitude $\phi_0 \lesssim \mpl$.
The constraints are stronger if non-renormalizable terms are taken
into account: 
\begin{equation}
V_{\rm NR} = \frac{|\lambda|^2}{\mpl^n} \phi^{4+n}.
\end{equation}
Non-renormalizable terms are unimportant for small enough masses,
$m_\phi \lesssim m_{\rm eff} = {V_{\rm eff}}''$.  For larger masses,
the curvaton slow-rolls in the non-renormalizable potential during and
after inflation.  In the post-inflationary epoch this leads to a huge
damping of the fluctuations, making it is impossible to obtain the
observed density contrast within the context of the curvaton
scenario~\cite{damping}.

The curvaton scenario should not alter the succesful predictions of
big bang nucleosynthesis (BBN).  This implies that the curvaton should
decay before the temperature drops below $\MeV$, and its coupling to
other fields cannot be arbitrarily small.  Gravitinos have only Planck
suppressed couplings and generically decay after BBN, thereby spoiling
BBN predictions if their number density is large.  To avoid gravitino
overproduction requires a reheat temperature $T_{\rm R} \lesssim 10^9
\GeV$: the inflaton should decay sufficiently late.  This also
constrains the curvaton scenario, since isocurvature perturbations are
converted in adiabatic perturbations only after inflaton decay. Note
that the entropy production at curvaton decay dilutes the gravitino
density, thereby ameliorating the gravitino problem.  In no-scale type
supergravity the gravitino mass is undetermined at tree level; the
gravitino problem is solved if the gravitino is heavy, $m_{3/2}
\gtrsim 100 \TeV$, and decays before BBN.

Finally, one should take into account various thermal effects. A large
thermal mass may be induced when the condensate is submerged in a
thermal bath. The heat bath is in thermal equilibrium for temperatures
$T \gtrsim h \phi$, where $h \phi $ is the effective mass of the
particles the curvaton couples to, and $h$ is the coupling constant.
Large thermal masses, $m_{\rm th} \gtrsim m_\phi$, induce early
oscillations.  The curvaton energy density not only decreases due to
the expansion of the universe, but also due to the decreasing mass:
$\rho_\phi(T) = \frac{m(T)}{m(T_0)} \frac{a(T_0)^3}{a(T)^3} \rho_\phi(T_0)$.
Further, it should be demanded that the curvaton does not decay too
early, through either thermal (through scattering) or resonant
decay~\cite{preheating}.  These constraints turn out to be less
stringent.

\section{Results}

\begin{center}
\begin{figure}
\hspace*{-0.9cm}
\psfig{figure=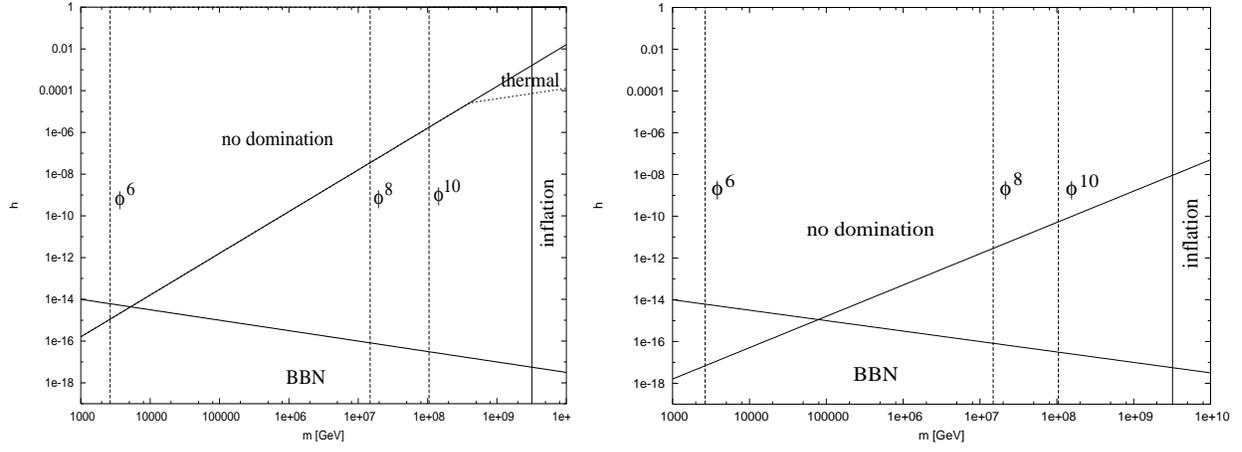,height=7cm}
\caption{Parameter space for curvaton domination ($f \gtrsim 0.5$).
In the plot on the left the reheating temperature is arbitrary high,
whereas in the plot on the right $T_{\rm R} \lesssim 10^9 \GeV$. The
constraints from BBN, domination, non-renormalizable terms,
$\phi$-dominated inflation, and thermal damping are shown.
\label{dom}}
\end{figure}
\end{center}

\begin{center}
\begin{figure}
  \hspace*{-1cm} \psfig{figure=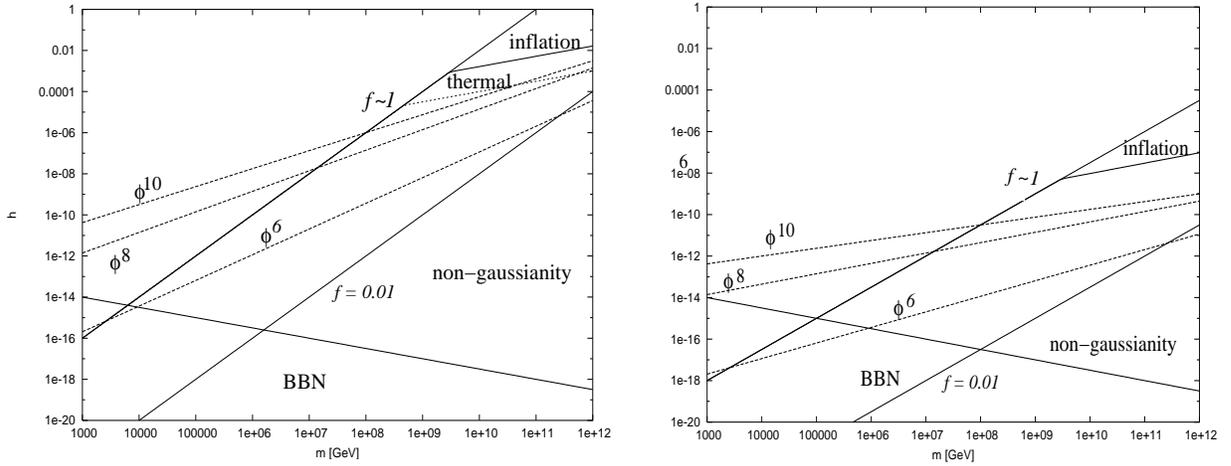,height=7.8cm}
\caption{Parameter space for curvaton non-domination ($10^{-2}
\lesssim f \lesssim 0.5$).  In the plot on the left the reheating
temperature is arbitrary high, whereas in the plot on the right
$T_{\rm R} \lesssim 10^9\GeV$. The constraints from BBN,
non-renormalizable terms, $\phi$-dominated inflation, non-Gaussianity,
and thermal damping are shown.
\label{nodom}}
\end{figure}
\end{center}

The parameter space for a succesfull curvaton scenario is shown in
Fig.~1 and Fig.~2.  In all plots $|\lambda| = 1$ and $\mpl = 1
/\sqrt{8\pi G}$.  We have assumed that the perturbations generated by
the inflaton are negligible small.

Fig.~1 shows the parameter space for {\it curvaton domination}; the
curvaton dominates the energy density for $f \gtrsim 0.5$.  In the
figure on the left the reheating temperature is arbritrary high,
whereas in the figure on the right the gravitino constraint is taken
into account and $T_{\rm R} \lesssim 10^9 \GeV$. In all parameter
space $m_\phi \sim 10^{-4} H_*$.  Models with $V_{\rm NR} \sim
\phi^{4+n}/\mpl^n$ and $n \leq 2$ are ruled out.  For higher values of
$n$, the curvaton scenario can be succesfull for curvaton masses in
the range $10^4 \GeV \lesssim m_\phi \lesssim 10^9 \GeV$. Couplings
have to be small $h \lesssim 10^{-6}$, even $h \lesssim 10^{-10}$ if
the gravitino constraint is taken into account, unless renormalizable
terms are absent to a very high order.  In all plots we assumed that
the curvaton has a typical initial value $\phi_0 \sim H_*^2/m_\phi$.
Dropping this assumption allows for larger coupling constants;
however, thermal damping should be taken into account for $h \gtrsim
10^{-5}$.

Fig.~2 shows the results for {\it non-domination}, $10^{-2} \lesssim f
\lesssim 0.5$; smaller values of $f$ lead to perturbations with
unacceptable large non-Gaussianity. In all parameter space $m_\phi
\sim 10^{-4} H_* / f$. The results are similar to the domination case:
a succesfull curvaton scenario needs small couplings, especially for a
low reheat temperature, and masses $m_\phi \gtrsim 10^4 \GeV$.  For
very large masses $m_\phi \to 10^{12} \GeV \approx 10^{-2} H_{\rm
max}$, larger couplings are possible.

The only constraints not considered yet pertain to residual
isocurvature perturbtations.  Isocurvature perturbations are defined
(on unperturbed hypersurfaces) as $S_i = 3 ({\mathcal R}_i - {\mathcal
R})$, with $i=$CDM, B for cold dark matter (CDM) and baryons
respectively. Further, ${\mathcal R}_i = -3H (\delta
\rho_i/\dot{\rho}_i) \propto \delta \rho_i /\rho_i$ is the curvature
perturbation of fluid $i$, and ${\mathcal R}$ the total curvature of
the universe.  The constraints are strongest for CDM.  There are three
possibilities~\cite{curv,gordon}:
\begin{itemize}
\item CDM number is created after curvaton decay.  The epoch of
creation is defined as the epoch after which the comoving CDM particle
number is conserved.  CDM and radiation have the same curvature
pertubation, and there are no residual isocurvature perturbations,
$S_{\rm CDM} = 0$.
\item CDM is created before curvaton decay.  If at creation the
curvaton energy density is still negligible small, $f \ll 1$, CDM has
a negligible curvature perturbation.  The isocurvature perturbation at
the epoch of last scattering then is $S_{\rm CDM} = - 3 {\mathcal R}$.
The isocurvature and curvature perturbations are correlated. The bound
from cosmic microwave background (CMB) measurements is $|S_{\rm
CDM}/{\mathcal R}| < 0.1$.  The constraint is weaker by a factor
$(\Omega_{\rm CDM} /\Omega_{\rm B}) \sim 10$ for baryons.  Creation of
dark matter well before curvaton decay is in conflict with experiment.
The same holds true for baryons, unless there is a cancelling CDM
isocurvature perturbation created by curvaton decay.
\item CDM is created by curvaton decay. Then $S_{\rm CDM} = 3 (
\frac{1-f}{f}) {\mathcal R}$.  The isocurvature and curvature
perturbations now are anti-correlated.  In this case the bound from
CMB measurements is $|S_{\rm CDM}/{\mathcal R}| < 0.2$.  CDM and
baryon isocurvature perturbations are unobservable small for $f> 0.9$
and $f>0.6$ respectively.
\end{itemize}

\section{Conclusions}
In the curvaton scenario, the adiabatic perturbations are not
generated by the inflaton field, but instead result from isocurvature
perturbations of some other field --- the curvaton field.  We have
analyzed various model independent constraints on such a scenario.
The curvaton scenario can work for small couplings $h \lesssim
10^{-6}$ and large masses $m_\phi \gtrsim 10^4 \GeV$.  Strong bounds
come from non-renormalizable operators in the potential, and from
gravitino overproduction.

One can ask whether there are any natural canditates for the curvaton.
Moduli and other fields with only Planck suppressed couplings
generically decay after big bang nucleosynthesis, thereby spoiling its
succesfull predicitions.  This problem is avoided for moduli with
large soft masses $m_{3/2} \gtrsim 100 \TeV$.  Fields parametrizing
flat directions in the potential of the supersymmetric standard model
typically have too small masses and too large couplings to play the
r\^ole of the curvaton.  Better curvaton candidates are the
right-handed sneutrino and the Peccei-Quinn axion, which can have
large masses and small couplings.  In all cases though, considerable
tuning of parameters is needed.  For example, for the right-handed
sneutrino curvaton, one needs to explain why the sneutrino mass is
much smaller than the grand unification scale.

\section*{Acknowledgments}
This work was supported by the European Union under the RTN contract
HPRN-CT-2000-00152 Supersymmetry in the Early Universe.


\section*{References}
\begin{thebibliography}{99}

\bibitem{cmb}
COBE:~http://aether.lbl.gov/www/projects/cobe/;
WMAP:~http://map.gsfc.nasa.gov/.

\bibitem{early}
S.~Mollerach,
\Journal{\PRD}{42}{313}{1990};
A.~D.~Linde and V.~Mukhanov,
\Journal{\PRD}{56}{535}{1997}.

\bibitem{curv}
K.~Enqvist and M.~S.~Sloth,
\Journal{\NPB}{626}{395}{2002};
D.~H.~Lyth and D.~Wands,
\Journal{\PLB}{524}{5}{2002};
D.~H.~Lyth, C.~Ungarelli and D.~Wands,
\Journal{\PRD}{67}{023503}{2003};
T.~Moroi and T.~Takahashi,
\Journal{\PLB}{522}{215}{2001}.

\bibitem{postma}
M.~Postma
\Journal{\PRD}{67}{063518}{2003}.

\bibitem{damping}
K.~Enqvist, A.~Jokinen, S.~Kasuya and A.~Mazumdar,
%``MSSM flat direction as a curvaton,''
arXiv:hep-ph/0303165.

\bibitem{preheating}
M.~Postma and A.~Mazumdar,
%``Resonant decay of flat directions: applications to curvaton scenarios, Affleck-Dine baryogenesis, and leptogenesis from a sneutrino
%condensate,''
arXiv:hep-ph/0304246.

\bibitem{gordon}
C.~Gordon and A.~Lewis,
%``Observational constraints on the curvaton model of inflation,''
arXiv:astro-ph/0212248.
%%CITATION = ASTRO-PH 0212248;%%


\end{thebibliography}
\end{document}